\documentclass[letterpaper,10pt,conference]{ieeeconf}
\IEEEoverridecommandlockouts

\usepackage{cite}
\usepackage{amsmath,amssymb,amsfonts}
\usepackage{dsfont}
\usepackage{graphicx}
\usepackage{booktabs}
\usepackage{xcolor}
\usepackage{subcaption}
\usepackage{url}
\usepackage{todonotes}

\newcommand{\R}{\mathbb{R}}
\newcommand{\E}{\mathbb{E}}

\newcommand{\Wass}{\mathbb{W}}
\newcommand{\norm}[1]{\left\lVert#1\right\rVert}

\newcommand{\CVaR}{\operatorname{CVaR}}

\usepackage[normalem]{ulem}

\newcommand{\myindexterms}[1]{%
\vspace{0.5em}
{\small\setlength{\baselineskip}{9pt}%
\noindent\textit{\textbf{Index Terms—}} #1\par}
}

\begin{document}

\title{Distributionally Robust Adaptive Iterative Covariance Steering for Small-Body  Proximity Operations}

\author{
Vivek Khatana$^{\dagger,\ddagger}$,
Aditya Gahlawat$^{\dagger}$,
Naira Hovakimyan$^{\dagger}$,
Petros Voulgaris$^{\ddagger}$\\
{\tt \small $^{\dagger}$\{vkhatana, gahlawat, nhovakim\}@illinois.edu} \\
{\tt \small $^{\ddagger}$\{vkhatana, pvoulgaris\}@unr.edu}\\
{\small $^{\dagger}$Department of Mechanical Science and Engineering,
University of Illinois Urbana-Champaign, Urbana, IL, USA }\\
{\small $^{\ddagger}$Department of Mechanical Engineering,
University of Nevada, Reno, NV, USA}
}

\maketitle

\begin{abstract}
Autonomous small-body proximity operations require continuous-thrust guidance under uncertain gravity, solar-radiation pressure (SRP), and stochastic disturbances. We combine iterative covariance steering (iCS) with $\mathcal{L}_1$ distributionally robust adaptive control ($\mathcal{L}_1$-DRAC) to provide risk-aware collision avoidance despite mismatch between the nominal planning model and the true dynamics. Iterative covariance steering plans a nominal feedback policy satisfying distributionally robust conditional value-at-risk keep-out constraints, while $\mathcal{L}_1$-DRAC compensates for dynamics uncertainty and bounds the residual distributional mismatch through a Wasserstein radius used by the planner. In $\mathbf{1000}$ paired Monte Carlo trials with gravity, SRP, and diffusion mismatch, $\mathcal{L}_1$-DRAC adaptive control reduces the keep-out violation rate from $\mathbf{32.7\%}$ to $\mathbf{0.2\%}$ and the physical collision rate from $\mathbf{16.4\%}$ to zero, without actuator saturation.
\end{abstract}

\myindexterms{\textbf{Space robotics, covariance steering, distributionally robust adaptive control, conditional value-at-risk, small-body proximity operations}}

\section{Introduction}

Small-body proximity guidance must accommodate weak nonlinear gravity, uncertain solar-radiation pressure (SRP), and strict collision-avoidance requirements. Covariance-steering methods have addressed stochastic small-body guidance using chance constraints and impulsive control~\cite{oguri2021,oguri2024,frassinella2026safe}, while sequential covariance steering has enabled continuous low-thrust guidance in cislunar dynamics~\cite{kumagai2025}. These approaches, however, rely on a specified planning model and can lose their safety guarantees under persistent dynamics mismatch. To maintain collision safety when the true dynamics differ from the
nominal planning model, we develop a risk-aware continuous-thrust
guidance framework combining iterative covariance steering (iCS),
$\mathcal{L}_1$ distributionally robust adaptive control
($\mathcal{L}_1$-DRAC), and Wasserstein distributionally robust
CVaR keep-out constraints~\cite{artzner1999coherent}. Unlike existing nominal covariance-steering
approaches that rely on a prescribed planning model, the proposed
architecture uses $\mathcal{L}_1$-DRAC to compensate for uncertain dynamics and to bound the residual state-distribution
mismatch. This bound directly defines the Wasserstein ambiguity
radius used to robustify the iCS safety constraint.

\section{Small-Body Stochastic Model}
Let the state be $X=[R^\top,V^\top]^\top\in\R^6$, where $R = [x,y,z]^\top \in \mathbb{R}^3$ and $V=[v_x,v_y,v_z]^\top \in \mathbb{R}^3$ denote position and velocity in a rotating asteroid-centered frame, and let $U\in\R^3$ be the commanded acceleration. A normalized augmented nonlinear Hill three-body problem (ANH3BP) model~\cite{frassinella2026safe} is: 
\begin{multline}\label{eq:true_sde}
     dX_t = \left[f_0(X_t,\beta)+B(U_t+H_\mu(X_t,t))\right] dt \\
     +\left[G_0+B H_\sigma(X_t,t)\right]dW_t, \quad X_0 = x_0,
\end{multline}
where $W_t$ is a Brownian motion, and
\begin{align*}
     f_0(X,\beta) =\begin{bmatrix}V \\ a_0(X, \beta)\end{bmatrix}, \ 
     B=\begin{bmatrix} 0_{3 \times 3} \\
     \mathbb{I}_3\end{bmatrix}, 
     G_0=\begin{bmatrix} 0_{3 \times 3} \\
     \sigma_0 \mathbb{I}_3\end{bmatrix}.
\end{align*}
Here, $\beta$ is the nominal SRP acceleration, $B \in \mathbb{R}^{6 \times 3}$ is the input operator, and 
\[ 
    a_0 (X,\beta) =\begin{bmatrix}
                     2v_y+3x-x/\norm{R}_2^3+\beta\\
                     -2v_x-y/\norm{R}_2^3\\
                     -z-z/\norm{R}_2^3 \end{bmatrix}.
\]
The function $f_0 \in \mathcal{C}\left(\mathbb{R}^6;\mathbb{R}^6\right)$, and the matrix $G_0 \in \mathbb{R}^{6\times 3}$ represent the \emph{known} components of the dynamics, whereas, the functions $H_\mu:\mathbb{R}^6\times[0,T]\rightarrow\mathbb{R}^3$ and
$H_\sigma:\mathbb{R}^6\times[0,T]\rightarrow\mathbb{R}^{3\times3}$ represent the \emph{uncertainties} in the drift and diffusion, respectively.

The iCS planner has access only to the nominal model obtained by setting $H_\mu=H_\sigma=0$ in~\eqref{eq:true_sde}
\begin{equation}\label{eq:nominal_sde}
    \begin{aligned}
        dX^\star_t = \left[f_0(X^\star_t,\beta)+BU^\star_t\right] dt + G_0 dW^\star_t, \ \ X^\star_0 = x^\star_0,
    \end{aligned}
\end{equation}
where $X^\star_t \in \mathbb{R}^6$ is the nominal state process and $W^\star_t \in \mathbb{R}^3$ is another Brownian motion
that is independent of $W_t$ driving~\eqref{eq:true_sde}. We denote the uncertain and nominal (known) probability laws of $X_t$ and $X^\star_t$ by $\mathbb{X}_t$ and $\mathbb{X}^\star_t$, respectively.


\subsection{Risk-Aware Small-Body Collision Avoidance}
Let $R_t$ be the spacecraft position relative to an asteroid. Let $r_{\rm col}=r_{\rm ast}+r_{\rm sc}$ denote the physical collision radius, where $r_{\rm ast}$ and
$r_{\rm sc}$ are the asteroid and spacecraft radii, respectively.
We define the keep-out radius as
\[
r_{\rm KO}=r_{\rm col}+d_{\rm safe},
\]
where $d_{\rm safe}>0$ is an additional mission-defined safety
buffer. The signed keep-out clearance is
\begin{equation}
D_t(R_t)=\lVert R_t\rVert_2-r_{\rm KO}.
\label{eq:clearance}
\end{equation}

Safety requires $D_t\geq0$. Equivalently, defining the safety loss as $L_t = -D_t$, we impose the following risk constraint:
\begin{equation}\label{eq:keepout_cvar}
    \operatorname{CVaR}_{\varepsilon_s}(L_t(R_t))\leq0,
    \qquad t\in[0,T],
\end{equation}
where, $\varepsilon_s\in(0,1)$ is the prescribed safety-risk tolerance, together with
$\mathbb E[X_T]=\mu_f$ and
$\operatorname{Cov}(X_T)\preceq\Sigma_f$.
The control problem is to minimize a statistical control cost subject to the uncertain dynamics~\eqref{eq:true_sde} and the safety~\eqref{eq:keepout_cvar} and terminal constraints.
To solve the posed problem, we employ iCS to compute a nominal trajectory and feedback policy satisfying the safety and terminal-distribution constraints. Because iCS relies on nominal dynamics, we augment it with $\mathcal{L}_1$-DRAC~\cite{L1DRAC} to compensate for model uncertainty and certify the residual distributional mismatch. Both components are described next.

\section{Distributionally Robust Adaptive iCS}

\subsection{Nominal Iterative Covariance Steering}

Since iCS is model-based~\cite{ridderhof2019nonlinear}, it requires a known nominal dynamics model~\eqref{eq:nominal_sde}. It repeatedly linearizes the nominal nonlinear dynamics about a reference trajectory and solves the resulting linear time-varying CS subproblems~\cite{liu2024optimal}, thereby extending finite-horizon CS to nonlinear stochastic systems~\cite{ridderhof2019nonlinear,kumagai2025}. At iCS iteration $j$, linearization of nominal dynamics~\eqref{eq:nominal_sde} about the current reference $(\overline{x}_t^{(j)} = [\overline{R}_t^{(j)}, \overline{V}_t^{(j)}]^\top,  \overline{u}_t^{(j)})$ gives
\begin{equation*}
 dX_t^\star = \left[ A_t^{(j)} X_t^\star + B_t^{(j)} U_t^\star + c_t^{(j)} \right]dt + G_t^{(j)} dW^\star_t,
\end{equation*}
where, 
\begin{align*}
    A_t^{(j)} &:= \frac{\partial f_0(x,\beta)}
    {\partial x}
    \bigg|_{x=\overline{x}^{(j)}_t}, \quad 
    B_t^{(j)}:= B, \\
    c_t^{(j)} &:= f_0(\overline{x}^{(j)}_t,\beta) - A_t^{(j)}\overline{x}^{(j)}_t - B_t^{(j)}\overline{u}_t^{(j)}, \quad  G_t^{(j)} := G_0.
\end{align*}

The iCS optimization problem is formulated with discrete-time dynamics. To this end, define $t_k := t_0 + k \Delta t$ and let $x^\star_k := X^\star_{t_k}$ and $u^\star_k := U^\star_{t_k}$ denote the sampled nominal state and
input, respectively. Under a zero-order-hold input, i.e., $U^\star_t = u^\star_k$ for $t \in [t_k, t_{k+1})$, the dynamics of $x^\star_k$  evolve as
\begin{align*}
    x^\star_{k+1} = A_{k,d}^{(j)} x^\star_k + B_{k,d}^{(j)} u^\star_k + c_{k,d}^{(j)} + G_{k,d}^{(j)} w^\star_k,
\end{align*}
where, $w_k \sim \mathcal{N}(0,\Sigma_{w_k}^{(j)})$. The matrices $A_{k,d}^{(j)} \in \mathbb{R}^{6 \times 6}, B_{k,d}^{(j)} \in \mathbb{R}^{6 \times 3}, G_{k,d}^{(j)}\in\mathbb R^{6\times3}$ and $\Sigma_{w_k}^{(j)} \in \mathbb{R}^{3 \times 3}$ are obtained from the analytic solution of~\eqref{eq:nominal_sde} under piecewise constant input $u^\star_k$ (see~\cite{sarkka2019applied}, Sections~6.1, 6.2,~\cite{ridderhof2019nonlinear}). The zero-order-held control policy is updated at time instants $t_k$ as
\begin{equation}\label{eq:discrete-policy}
 u_k^\star := v^\star_k+K^\star_k(x_k^\star-\mu_k^\star),                 
\end{equation}
where $\mu_k^\star:=\E[x_k^\star]$. The feedforward and feedback gains $(v^\star_k, K^\star_k)$ are determined by solving an optimization problem with a statistical control cost function (see problem~{(34)} in~\cite{liu2024optimal} for a formulation).

\subsection{Wasserstein DR-CVaR Safety}
At reference position $\bar R_k^{(j)}$, let $n_k:=\bar R_k^{(j)}/\|\bar R_k^{(j)}\|_2$. The supporting half-space gives
the conservative loss
\[
    \widetilde L_k(R_k)=r_{\rm KO}-n_k^\top R_k.
\]

Therefore, a conservative nominal safety constraint is 
\begin{equation}\label{eq:nominal_CVAR}
   \hspace{-0.1in} \CVaR_{\varepsilon_k} (\widetilde L_k(R_k))\leq 0, \ k \in \left \{0,\Delta t, \dots, N := \left \lceil \frac{T}{\Delta t} \right \rceil \right \}
\end{equation}
where, $\widetilde L_k(R_k) = - \widetilde D_k(R_k)$, $\sum_{k=1}^{N} \varepsilon_k = \varepsilon_s$. The iCS planner predicts a nominal Gaussian state distribution $ x_k^\star \sim \mathcal{N}(\mu_k^\star,\Sigma_k^\star)$
where $\mu_{R,k}^\star$ and $\Sigma_{RR,k}^\star$ are the position components of the
predicted mean and covariance. Under this distribution, loss $L_k(R_k)$ is Gaussian with mean and standard deviation 
\[
    \mu_{L,k} := r_{\rm KO}-n_k^\top\mu_{R,k}^\star, \quad \sigma_{L,k} := \sqrt{n_k^\top\Sigma_{RR,k}^\star n_k}.
\]
The constraint~\eqref{eq:nominal_CVAR} becomes
\begin{equation}\label{eq:gaussian_keepout_cvar}
    r_{\rm KO} - n_k^\top\mu_{R,k}^\star
    + \kappa_{\varepsilon_k}
    \sqrt{n_k^\top\Sigma_{RR,k}^\star n_k}
    \leq 0,
\end{equation}
where $\kappa_{\varepsilon_k} := \frac{
        \phi \left(\Phi^{-1}(1-\varepsilon_k)\right)
    }{\varepsilon_k}$ where $\phi$ and $\Phi$ are the standard-normal probability density and cumulative distribution functions, respectively. 

The constraint \eqref{eq:gaussian_keepout_cvar} is valid when the predicted
Gaussian distribution $\mathbb{X}_k^\star := \mathcal{N}(\mu_k^\star, \Sigma_k^\star)$ accurately represents the distribution of the true state $X_k$ of the spacecraft.
Uncertain gravity, SRP, nonlinear propagation, and residual adaptive-control error can cause the true distribution of $X_k$ to differ from $\mathbb{X}_k^\star$. We account for this discrepancy by requiring the true state distribution to lie in the Wasserstein distributional ambiguity set $\mathbb{B}_{\rho}(\mathbb{X}_k^\star) = \left\{
        \mathbb{Q}: \Wass_2(\mathbb{Q},\mathbb{X}_k^\star)\leq\rho \right\}$,
where $\Wass_2$ is the $2$-Wasserstein distance and $\rho$ is a bound on the state distribution mismatch. The distributionally robust safety requirement is
\begin{equation}
    \textstyle \sup_{\mathbb{Q}\in\mathbb{B}_{\rho}(\mathbb{X}_k^\star)}
    \CVaR_{\varepsilon_k}^{\mathbb{Q}}
    \left(\widetilde L_k(R_k)\right)
    \leq0.
    \label{eq:dr_cvar_requirement}
\end{equation}
Thus, the safety constraint must hold not only for the nominal Gaussian distribution but for every state distribution within Wasserstein distance $\rho$ of $\mathbb{X}_k^\star$. With~\eqref{eq:gaussian_keepout_cvar}, a tractable sufficient condition for~\eqref{eq:dr_cvar_requirement} is~\cite{L1DRCS, zhang2025short}
\begin{equation} \label{eq:dr_cvar_keepout}
    \textstyle r_{\rm KO} - n_k^\top\mu_{R,k}^\star + \kappa_{\varepsilon_k}
    \sqrt{n_k^\top\Sigma_{RR,k}^\star n_k}
    + \left(\displaystyle \frac{\rho}{\sqrt{\varepsilon_k}} \right)
    \leq 0.
\end{equation}
The radius $\rho$ of the Wasserstein ambiguity set in~\eqref{eq:dr_cvar_keepout} is provided by the $\mathcal{L}_1$-DRAC controller described next.

\subsection{$\mathcal{L}_1$-DRAC Augmentation}

The uncertain system receives 
\[
    U_t=U_t^\star+U_{\mathcal L_1,t},
\]
where $U_t^\star$ is determined by iCS and $U_{\mathcal L_1,t}$ is the adaptive control input generated by the predictor, piecewise-constant adaptation law, and low-pass filter of~\cite{L1DRAC}. Under its stated regularity and stability conditions, $\mathcal{L}_1$-DRAC bounds the Wasserstein distance between the nominal and true state distributions by 
\[
    \mathbb{W}(\mathbb X_t,\mathbb X_t^\star)\leq\rho \ \forall t \in [0,T] \implies \mathbb{X}_t \in \mathbb{B}_{\rho}(\mathbb{X}_t^\star) \ \forall t \in [0,T].
\]
Thus, $\mathcal{L}_1$-DRAC adaptation compensates model uncertainty, and $\rho$ accounts for the residual distributional mismatch in~\eqref{eq:dr_cvar_keepout}.

\section{Simulation Study} \label{sec:simulation-study}
We consider a representative half-revolution proximity maneuver governed by
the normalized ANH3BP with $\beta=0.01$. The asteroid and spacecraft radii are $r_{\rm ast}=0.185$~km and $r_{\rm sc}=0.0037$~km, giving the physical collision radius
$r_{\rm col}=0.1887$~km, respectively. With the additional safety buffer $d_{\rm safe}=0.001$~km, the keep-out radius is $r_{\rm KO}=r_{\rm col}+d_{\rm safe}=0.1897$~km. Using the length scale $L=0.185$~km, the corresponding normalized radii are $r_{\rm col}/L=1.02$ and $r_{\rm KO}/L=1.0254$. The reference radius
is $1.1$, the normalized maneuver horizon is $T=4$, and $ \mu_0=[0,-1.1,0,0.86,0,0]^\top, \mu_f=[0,1.1,0,-0.86,0,0]^\top.$ 
We use $\Sigma_0=10^{-4}\operatorname{diag}(8,8,8,40,40,40)$,
$\Sigma_f=10^{-2}\mathbb{I}_6$, $\Delta t=0.1$ for iCS, and $\varepsilon_s =0.01$.

\subsection{System Uncertainty} The true system~\eqref{eq:true_sde} contains uncertainties that are not included in the nominal iCS planning dynamics. In particular, true dynamics include uncertainty in the asteroid gravitational parameter and in the solar-radiation-pressure acceleration. The corresponding acceleration perturbation is 
\begin{equation*}
    \Delta a_{\mu}(X) = -\delta_{\mu}\frac{R}{\lVert R\rVert_2^3}
    +\delta_{\beta}e_x, \quad e_x = [1, 0, 0]^\top.
\end{equation*}
Therefore, the matched drift-uncertainty function is 
    \[  
    H_{\mu}(X)
    = -\delta_{\mu}\frac{R}{\lVert R\rVert_2^3}
    +\delta_{\beta}e_x.
    \]
The gravity error $\delta_{\mu}$ represents a fractional error in the normalized asteroid gravitational parameter. In each independent system run it is sampled according to $\delta_{\mu}
    \sim
    \operatorname{clip}
    \left(
        \mathcal{N}(0,0.85^2),
        -0.99,
        0.99
    \right)$. The resulting true gravitational scaling is $1+\delta_{\mu}$. The lower clipping limit ensures that this scale remains positive. The solar-radiation-pressure error is sampled as $\delta_{\beta}
    \sim
    \operatorname{clip}
    \left(
        \mathcal{N}(0,0.03^2),
        -0.08,
        0.08
    \right)$ at each system run. Additionally, we use a constant matched diffusion-uncertainty function $H_{\sigma}(X)=0.30 \mathbb{I}_3$. These uncertainty levels are intended to provide a representative stress test of persistent gravity, SRP, and diffusion mismatch rather than a mission-calibrated uncertainty model for a specific asteroid. Combining the nominal dynamics with the matched drift and diffusion uncertainties gives the true model    
\begin{equation*}
\begin{aligned}
dX_t= & 
\left[
f_0(X_t,\beta)+
B\left(
U_t-\delta_\mu\frac{R_t}{\|R_t\|_2^3}
+\delta_\beta e_x
\right)
\right]dt\\
&+
\left(G_0+ B 0.30\mathbb{I}_3\right)dW_t .
\end{aligned}
\end{equation*}

\subsection{Control Policy Implementation} 
The control applied to the uncertain system is 
\[
u_{\mathrm{app}}(t) = \operatorname{clip} \left(
        U_t^\star + U_{\mathcal{L}_1,t},
        -20,
       20
    \right),
\]
where $U_t^\star$ is the iCS control policy in~\eqref{eq:discrete-policy}, and $U_{\mathcal{L}_1,t}$ is the
$\mathcal{L}_1$-DRAC control input. The $\mathcal{L}_1$-DRAC parameters are selected as $\{ \omega=20,T_s=10^{-3},\lambda_s=50\}$ leading to a corresponding Wasserstein ambiguity set radius $\rho = 0.128$. The average iCS solution time was $10.85$~s. This solve time corresponds to the high-level iCS planning update, rather than the lower-level adaptive-control update; reducing planning latency through warm starts and more efficient sequential implementations is an important direction for real-time deployment.

\begin{figure}[b]
    \centering
    \includegraphics[width=\linewidth,trim={0 0 0 0},clip]
    {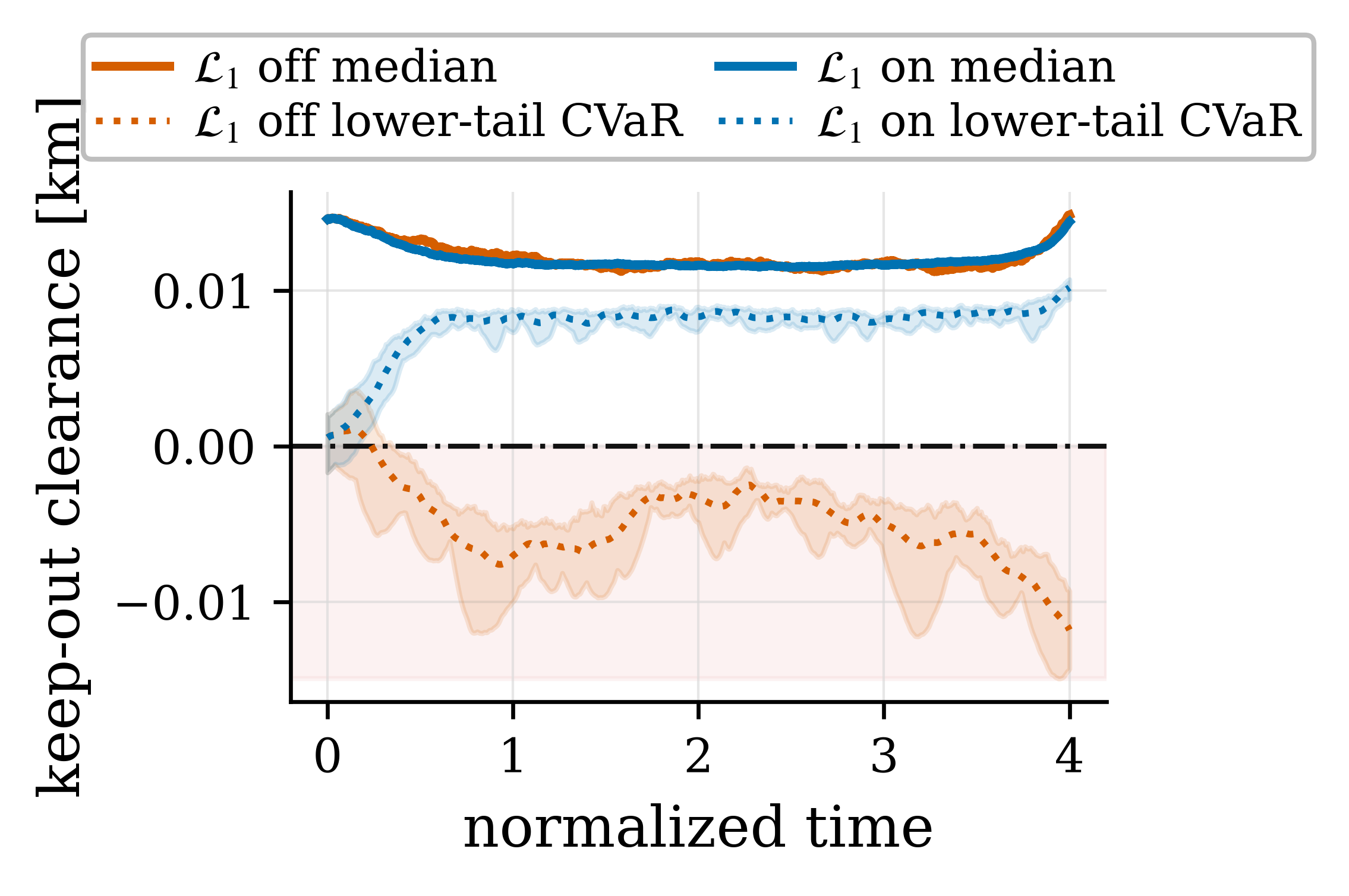}
    \caption{\small Signed keep-out clearance over the maneuver. Solid curves show the ensemble medians, shaded regions span the empirical $0^{th}$--$1^{st}$ percentile clearance tails, and dotted curves show the corresponding lower-tail CVaR, $\operatorname{CVaR}_{0.01}(D_t)$. The dash-dotted horizontal line marks the keep-out boundary; negative clearance denotes a keep-out violation.}
    \label{fig:clearance}
\end{figure} 

\vspace{0.1in}
\section{Simulation Results}

\begin{figure*}[t]
\centering

\begin{subfigure}[t]{0.4\textwidth}
    \centering
\includegraphics[width=\linewidth,trim={0 0 0 0},clip]
    {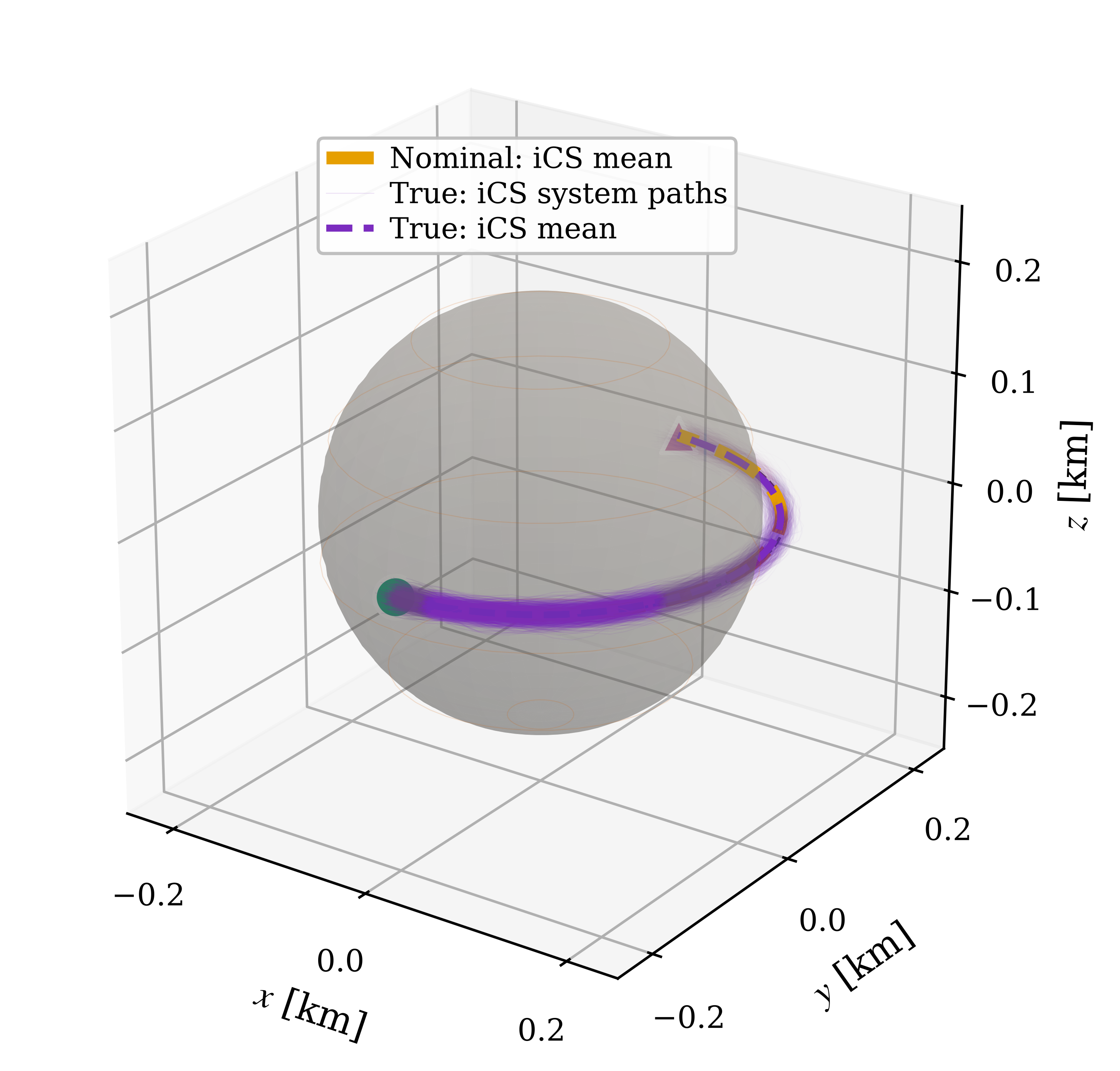}
    \caption{\small True-system trajectories: $\mathcal L_1$-DRAC off.}
    \label{fig:trajectory-3d-off}
\end{subfigure}
\begin{subfigure}[t]{0.4\textwidth}
    \centering
\includegraphics[width=\linewidth,trim={0 0 0 0},clip]
    {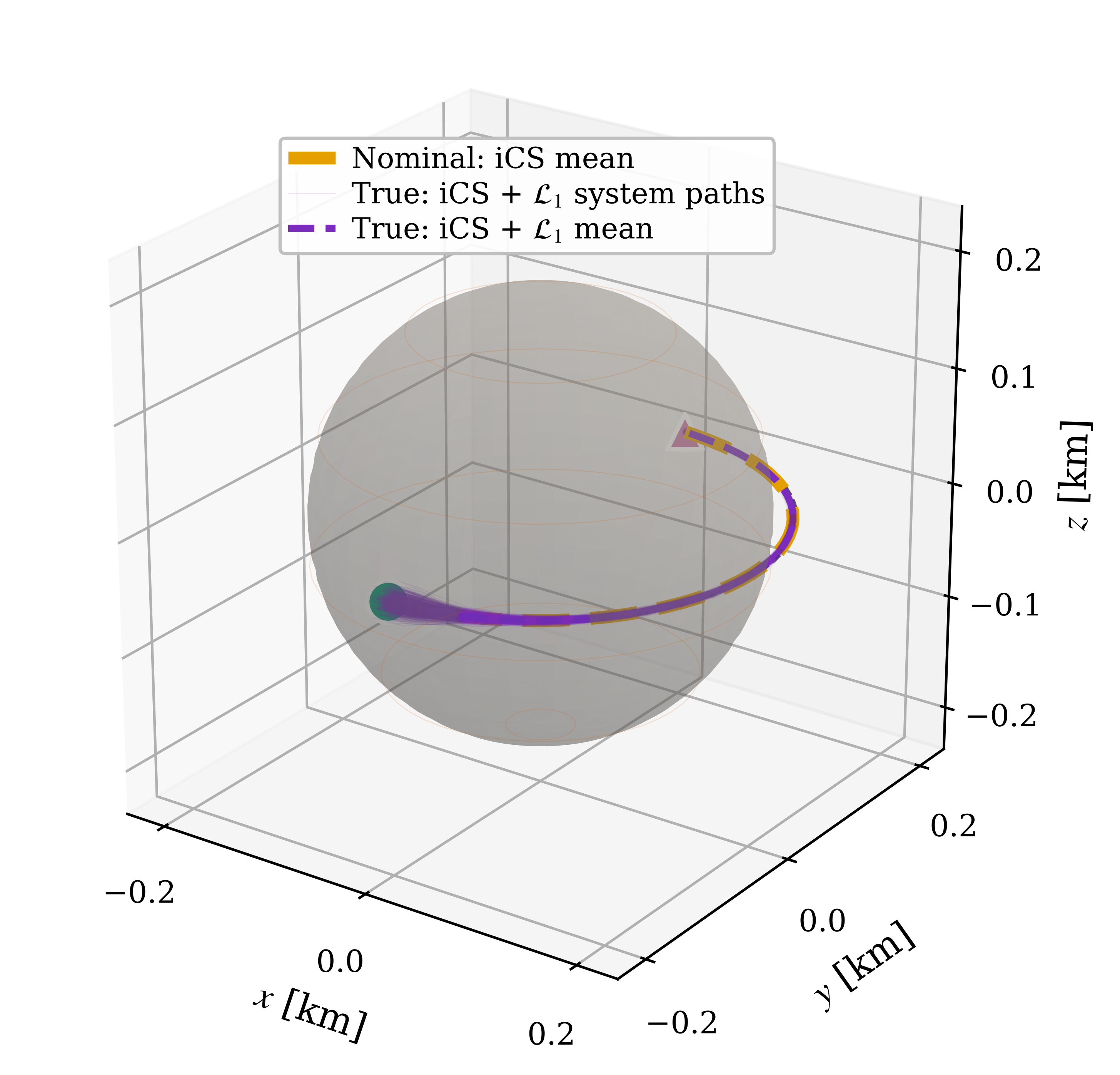}
    \caption{\small True-system trajectories: $\mathcal L_1$-DRAC on.}
    \label{fig:trajectory-3d-on}
\end{subfigure}

\vspace{0.3em}

\begin{subfigure}[t]{0.4\textwidth}
    \centering
    \includegraphics[width=0.82\linewidth,trim={0 0 0 0},clip]
    {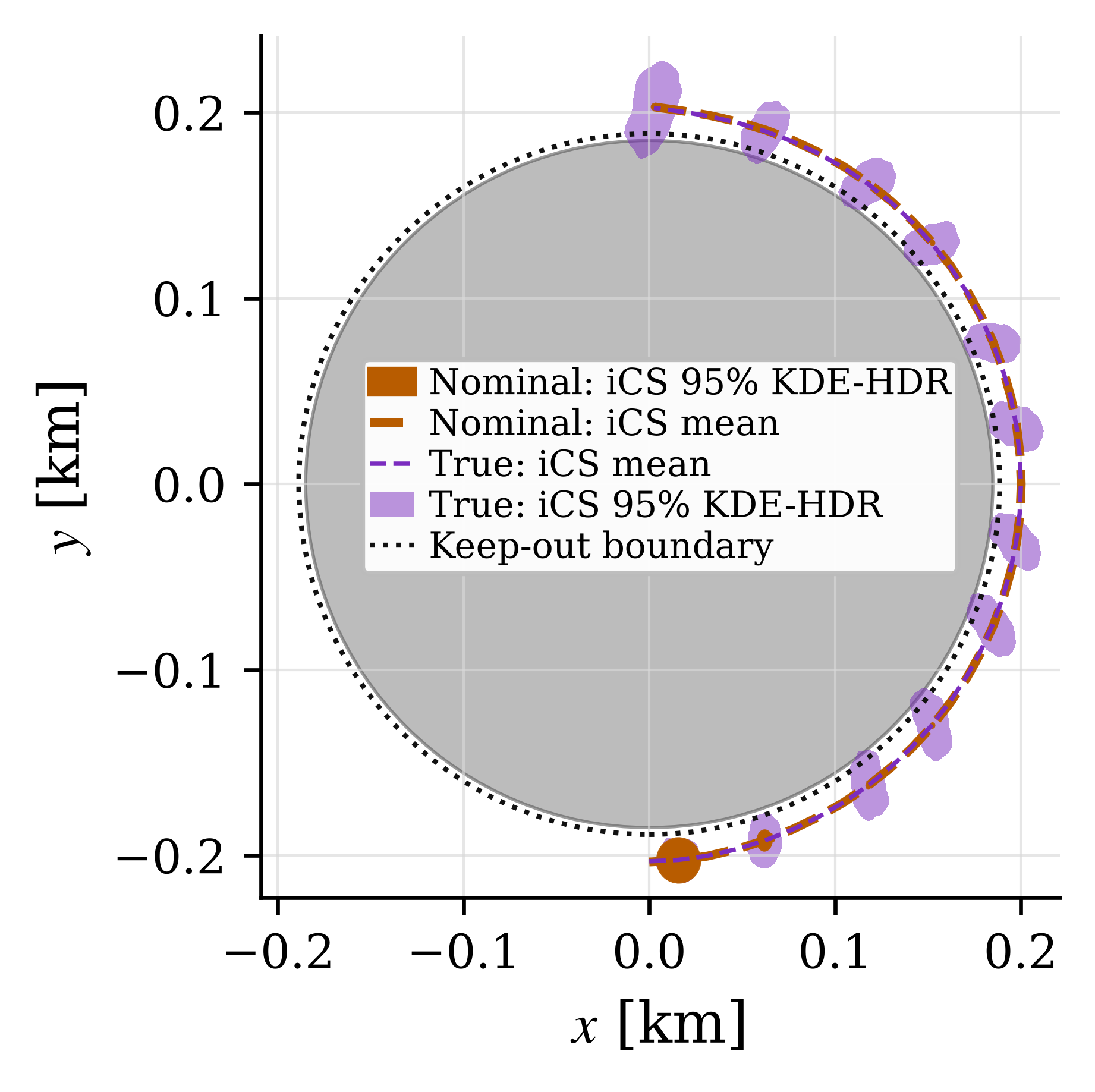}
    \caption{\small Nominal and empirical distributions: $\mathcal L_1$-DRAC off.}
    \label{fig:distribution-2d-off}
\end{subfigure}
\begin{subfigure}[t]{0.4\textwidth}
    \centering
    \includegraphics[width=0.82\linewidth,trim={0 0 0 0},clip]
    {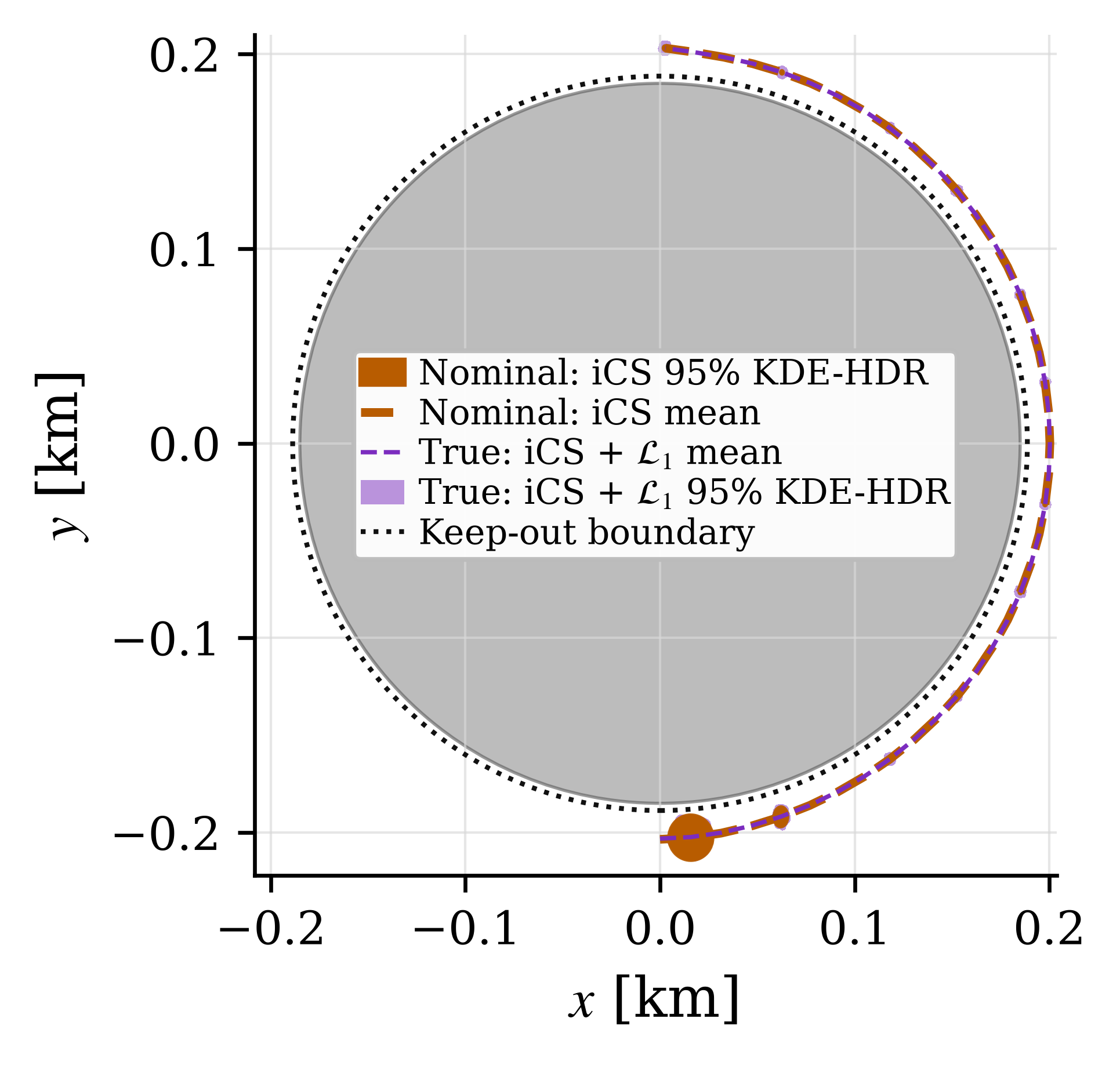}
    \caption{\small Nominal and empirical distributions: $\mathcal L_1$-DRAC on.}
    \label{fig:distribution-2d-on}
\end{subfigure}
\caption{\small Effect of $\mathcal L_1$-DRAC on the uncertain
closed-loop spacecraft motion. Panels (a)--(b) show representative
true-system trajectories and their empirical means under the same
nominal iCS policy with adaptation disabled and enabled, respectively.
Panels (c)--(d) compare the nominal and empirical means and their
$95\%$ kernel-density-estimated highest-density regions (KDE-HDRs).
The gray body represents the asteroid, and the dotted circle in
(c)--(d) denotes the keep-out boundary. The green circle and magenta
triangle in (a)--(b) mark the initial and terminal positions.
All distributional statistics are computed using the complete
$1000$-trial ensemble.}
\label{fig:trajectory-distribution-comparison}
\end{figure*}

We evaluate the safety and robustness of the closed-loop architecture by comparing nominal iCS with and without the adaptive $\mathcal{L}_1$-DRAC augmentation under
persistent dynamics mismatch. Both cases use the same nominal iCS policy and paired uncertainty realizations; they differ only in whether $U_{\mathcal L_1,t}$ is applied. Importantly, $U_t^\star$ in both cases is the closed-loop iCS policy in~\eqref{eq:discrete-policy}, including the optimized feedback gain $K_k^\star$; thus the comparison isolates the effect of the adaptive augmentation rather than the presence of nominal state feedback. A keep-out violation is recorded if $\min_t(\|R_t\|_2-r_{\rm KO})<0$, whereas a physical collision is recorded if $\min_t(\|R_t\|_2-r_{\rm col})<0$.

Across $1000$ paired Monte Carlo trials, iCS without adaptation produced $327$ keep-out violations and $164$ physical collisions. With $\mathcal{L}_1$-DRAC, these counts decreased to two keep-out violations and no collisions, reducing the respective empirical rates from $32.7\%$ to $0.2\%$ and from $16.4\%$ to zero. No actuator saturation was observed.

Figure~\ref{fig:clearance} examines the safety-critical lower tail of the clearance distribution. Although the two controllers produce similar median clearances, their tail behavior differs substantially. Without adaptation, the worst $1\%$ clearance tail extends deeply into the unsafe region, and its lower-tail CVaR becomes negative over most of the maneuver. With $\mathcal L_1$-DRAC, the tail contracts toward the median and the lower-tail CVaR remains nonnegative over the controlled horizon. The two adaptive violations appear within the displayed worst-$1\%$ tail but do not make its average negative, consistent with the prescribed CVaR risk tolerance.

Figure~\ref{fig:trajectory-distribution-comparison} compares the resulting trajectory ensembles and position distributions. Without adaptation, the true-system trajectories exhibit substantial dispersion relative to the nominal iCS trajectory (Fig.~\ref{fig:trajectory-3d-off}). Consequently, the empirical mean and $95\%$ kernel-density-estimated highest-density regions (KDE-HDR) depart from the nominal distribution and extend toward the keep-out region (Fig.~\ref{fig:distribution-2d-off}). With $\mathcal L_1$-DRAC, the trajectories remain concentrated around the planned path (Fig.~\ref{fig:trajectory-3d-on}), and the empirical mean and KDE-HDR closely track their nominal counterparts (Fig.~\ref{fig:distribution-2d-on}).

\section{Conclusion}
This paper developed an adaptive, distributionally robust covariance steering framework for continuous-thrust small-body proximity operations. iCS plans a safe state distribution using the nominal dynamics, while $\mathcal L_1$-DRAC compensates for matched acceleration uncertainty. The distributional mismatch is captured by a  Wasserstein ambiguity set in the DR-CVaR keep-out constraints. In $1000$ paired Monte Carlo trials, adaptation reduced the keep-out violation rate from $32.7\%$ to $0.2\%$ and eliminated physical collisions. Future work will assess sensitivity across additional uncertainty models and small-body scenarios, reduce iCS planning latency, and validate the framework using higher-fidelity and hardware-in-the-loop simulations.

\section*{{Acknowledgment}}
The authors used ChatGPT (OpenAI) to edit the manuscript. The authors independently conceived all research questions, designed architectures, algorithms, experimental methods, and performed all validations, interpretations, and scientific conclusions.

\bibliographystyle{IEEEtran}
\bibliography{References}

\end{document}